%% file: Template.tex
\documentclass[10pt]{article}
\usepackage{spconf,graphicx}

\title{Kalman Filter based MIMO CSI Phase Recovery for COTS WiFi Devices}
\name{Chu Li*
	, Jeremy Brauer**, Aydin Sezgin*, Christian Zenger**}
\address{*Ruhr-Universit\"at Bochum, **PHYSEC GmbH\\
	Email:  \{chu.li, aydin.sezgin\}@rub.de, 
	\{jeremy.brauer, christian.zenger\}@physec.de} 
\input{preamble}

\begin{document}

	
	%
	%
	%

	%
	\maketitle
	
	%
	\begin{abstract}
		Recently channel state information (CSI) measurements from commercial multi-input multi-output (MIMO) WiFi systems have been ubiquitously used for different wireless sensing applications. However, the phase of the CSI realizations is usually distorted severely by phase errors due to the hardware impairments, which significantly reduce the sensing performance. 
		In this paper, we directly utilize the modeling of the phase distortions caused by the hardware impairments and propose an adaptive CSI estimation approach based on Kalman filter (KF) with maximum-a-posteriori (MAP) estimation that considers the CSI from the previous time. The performance of the proposed algorithm is compared against the Cramér–Rao lower bound (CRLB). Simulation and experimental results demonstrate that our approach can track the channel variations while eliminating the phase errors accurately. 
	\end{abstract}
	\begin{figure*}[b]
		\vspace{-0.3cm}
		\parbox{\hsize}{\em
%
© 20XX IEEE.  Personal use of this material is permitted.  Permission from IEEE must be obtained for all other uses, in any current or future media, including reprinting/republishing this material for advertising or promotional purposes, creating new collective works, for resale or redistribution to servers or lists, or reuse of any copyrighted component of this work in other works.
	}\end{figure*}
	%
	%
	\vspace{-0.5cm}
	\section{Introduction}
	\label{sec:intro}
	
	In recent years the CSI can be obtained from commercial off-the-shelf (COTS) WiFi devices, which greatly facilitates the use of ubiquitous WiFi signals in various indoor WiFi sensing applications, such as positioning \cite{tadayon2019decimeter}, human identification/authentication \cite{chen2017rapid}, activity recognition \cite{arshad2017wi}, etc. In contrast to conventional sensor-based and video-based solutions WiFi signal based sensing systems are easy to implement and has lower costs.
	
	\noindent However, raw CSI measurements from the COTS WiFi devices contain random phase errors caused by the imperfect synchronization between the transmitter and receiver. The time shift from the packet boundary detection results in packet detection delay (PDD), which leads to random phase \emph{slope} error. The second important contributor is the carrier frequency offset (CFO). Although a CFO corrector is defined in the WiFi standard, the compensation is still incomplete. The residual CFO leads to random phase \emph{offset} error~\cite{ RCFOcompension,zhu2018splicer}. 
	In many recent CSI-based user authentication studies this problem is avoided by completely ignoring the CSI phase and focusing only on the CSI magnitude~\cite{liu2018authenticating}. The authors in~\cite{ RCFOcompension} proposed a multi-scale sparse recovery algorithm to first estimate the PDD and then compensate the residual CFO by the Multiple Signal Classification (MUSIC) algorithm. However, in order to get better estimation accuracy, multiple CSI realizations are required. Moreover, the multi-scale sparse recovery needs to be performed at every received packet, resulting in high computational complexity. Some other studies attempt to explicitly eliminate the random phase distortions. A conventionally used method is to perform linear regression on the raw measured CSI phase~\cite{kotaru2015spotfi,ma2018signfi,ma2019wifi}. However, the estimated phase based on regression includes both the phase distortions and the true CSI phase that cannot be separated efficiently.
	
	\noindent To address these problems, we utilize an adaptive KF based algorithm to recover the CSI from the distorted observations. In contrast to the algorithm in~\cite{ RCFOcompension}, which depends on multiple CSI realizations, our proposed algorithm only needs the knowledge of the previous and current CSI. Since the COTS WiFi systems usually have multiple antennas, in this work we consider MIMO setups and extend our previous work\cite{chu2019phase} in which we considered a single-input single-output (SISO) system. Moreover, to evaluate the performance of our algorithm, we perform Monte Carlo simulations and compare the mean square error (MSE) with the CRLB. In addition to the analysis of the algorithm with synthetic data, CSI data from COTS WiFi devices are used to verify the validity of the proposed algorithm with experimental data. 
	
	\noindent The rest of the paper is organized as follows. The system model is given in Sec.~\ref{sec:system}. We introduce the adaptive KF algorithm with MAP in Sec.~\ref{sec:algorithm}. The CRLB for the MAP estimator and the Kalman filtering based estimator are derived in Sec.~\ref{sec:CRLB}. Simulation and experimental results are given in Sec.~\ref{sec:results}. Finally, Sec.~\ref{sec:conclusion} concludes the paper. 

	\section{SYSTEM MODEL}
	\label{sec:system}
	In this work, we consider a MIMO system with $N_T$ transmit antennas sending WiFi packets to $N_R$ receive antennas.  Let $N =N_T \cdot N_R$ denote the number of channels and $M$ denote the discrete Fourier transform (DFT) size. We observe $Q\leq M$ subcarriers used for the channel estimation. The observation model at time $k$ is defined as
	\vspace{-0.2cm}
	\begin{align}
	\label{eq:obsEq}
	\mybold{H}_{\text{Obs},k} = e^{j\Omega_{0,k}} \mybold{E}(\Omega_{d,k}) \mybold{C}\mybold{H}_{k}+\mybold{W}_{k},
	\end{align}
	in which
	$\mybold{H}_{\text{Obs},k} = 
	[\mybold{h}_{\text{Obs},k}^{1}, \mybold{h}_{\text{Obs},k}^{2},  \ldots,  \mybold{h}_{\text{Obs},k}^{N} ]
	$ is a $Q\times N$ observed CSI matrix in the DFT domain,
	$\Omega_{0,k}$ is the phase offset error. The diagonal matrix $\mybold{E}_k$ represents the phase slope error, which can be written as \vspace{-0.2cm}
	\begin{align}
	\label{eq:diagonalPhase}
	\mybold{E} = 
	\diag
	\begin{bmatrix}
	e^{j\Omega_{d,k}q_1}, & e^{j\Omega_{d,k}q_2}, & \ldots, & e^{j\Omega_{d,k}q_Q}  
	\end{bmatrix},
	\end{align}
	where $q_1,q_2,\ldots,q_Q\in\mathcal{Q}$ denote the pilots indices. Let $L$ be the channel length. The $Q \times L$ matrix $C$ represents the DFT matrix with ${[C]}_{m,l}=e^{-j\frac{2\pi }{M}\cdot q_m\cdot l}$. The $L \times N$ matrix 
	$\mybold{H}_{k} = [
	\mybold{h}_{k}^{1},  \mybold{h}_{k}^{2},  \ldots,  \mybold{h}_{k}^{N}
	]$ in~\eqref{eq:obsEq} represents the MIMO channel in the time domain. The $Q\times N$ matrix $\mybold{W}_{k}$ is parallel samples of independent circularly-symmetric Gaussian white noise with covariance $\expv\left[\mybold{\omega}_{k}^{i}\mybold{\omega}_{k}^{iH}\right]=\sigma^{2}_{w}$, where $\mybold{\omega}_{k}^{i}$ is the $i$-th column of $\mybold{W}_{k}$.
	We model the time-varying nature of the channel by a first order auto-regressive (AR1) model
	\vspace{-0.3cm}
	\begin{align}
	\label{eq:sysEq}
	\mybold{H}_{k}= \alpha\mybold{H}_{k-1}+\mybold{V}_{k},
	\end{align}
	where $\mybold{V}_{k} = 
	[
	\mybold{v}_{k}^{1},  \mybold{v}_{k}^{2},  \ldots,  \mybold{v}_{k}^{N}
	]$ denotes the complex white Gaussian process noise with $\expv\left[\mybold{v}_{k}^{i}\mybold{v}_{k}^{iH}\right] = \sigma^{2}_{v(i,k)}\eye{L}$, $\alpha $ is the transition parameter that represents the similarity between the channel in the previous time $k-1$ and the current time $k$. In this paper, we assume that the channel variations, namely the parameter $\alpha$ is the same for all channels of the MIMO system.

	\section{Algorithm}
	\label{sec:algorithm}
	
	To recover the CSI from the distorted observations we propose an adaptive estimation algorithm based on KF and MAP. A KF is an efficient algorithm that can estimate the internal hidden state of a linear dynamic system from a series of noisy measurements~\cite{zarchan2000fundamentals}. In this paper, we consider the channel in the time domain $\mybold{H}_k$ as the internal hidden state and then recover it from the distorted observation $\mybold{H}_{Obs,k}$. Based on the state-space model defined in~\eqref{eq:obsEq} and~\eqref{eq:sysEq} we derive the adaptive Kalman process with MAP as follows.
	
	\noindent  Let $\mybold{P}_{k|k-1}=[
	\mybold{p}_{k|k-1}^{1},  \mybold{p}_{k|k-1}^{2},  \ldots,  \mybold{p}_{k|k-1}^{N}]
	$ denote the predicted estimation error covariance and  $\mybold{P}_{k-1|k-1}=[
	\mybold{p}_{k|k}^{1},  \mybold{p}_{k|k}^{2}, \ldots,  \mybold{p}_{k|k}^{N}
	] $ denote the updated estimation error covariance at time $k$, respectively. Furthermore, we use $\hat{\mybold{H}}_{k|k-1}$ and $\hat{\mybold{H}}_{k|k}$ to denote the predicted state and the updated state at time $k$. We first initialize the channel state $\hat{\mybold{H}}_{0|0} = \mybold{0}$, the estimation error covariance $\mybold{P}_{0|0} = \mybold{1}$. 
	The prediction step of the KF process is given by \vspace{-0.3cm}
	\begin{align}
	\hat{\mybold{H}}_{k|k-1}=\alpha \hat{\mybold{H}}_{k-1|k-1},
	\end{align} 
	\vspace{-0.3cm}
	\begin{align}
	\label{eq:ppred}
	\mybold{P}_{k|k-1} = {\alpha}^2 \mybold{P}_{k-1|k-1} +\mybold{R}_{k},
	\end{align} where $\mybold{R}_{k} =  [\mybold{r}^{1}_{k},  \mybold{r}^{2}_{k},   \ldots,  \mybold{r}^{N}_{k}] $ represents the covariance of the process noise with $ \mybold{r}^{i}_{k} = \sigma^{2}_{v(i,k)} {\mathds{1}}_L $. After the prediction we apply the MAP estimation to obtain the phase distortion parameters $\Omega_{d,k}$ and $\Omega_{0,k}$. For simplicity the MIMO channels are assumed to be uncorrelated. The joint probability density of $\hat{\mybold{H}}_{k|k-1}$ and $\mybold{H}_{\text{Obs},k}$ is assumed to follow a parallel complex Gaussian distribution. We use $f_{\mybold{H}_{\text{Obs}}}$ to denote the conditional density function of $\mybold{H}_{\text{Obs}}$ given the predicted channel state $\hat{\mybold{H}}_{k|k-1}$, the phase distortion parameters $\Omega_{d,k}$ and $\Omega_{0,k}$. 
	$\Omega_{d,k}$ and $\Omega_{0,k}$ are assumed to be independent and uniformly distributed. Thus, the MAP estimator is equivalent to maximum likelihood (ML). 
	Here, we introduce $\mybold{B}_k= e^{j\Omega_{0,k}} \mybold{E}(\Omega_{d,k}) \mybold{C}$ for convenience. The negative log-likelihood (NLL) function at time $k$ is given by \vspace{-0.3cm}
	\begin{align}
	\label{eq:loglikelihood}
	&g\left(\Omega_{d,k}, \Omega_{0,k}\right) \nonumber\\
	&\quad :=-\ln f_{\mybold{H}_{\text{Obs}}}\left( \mybold{H}_{\text{Obs},k} \Big| \hat{\mybold{H}}_{k|k-1},{\Omega}_{0,k}, \Omega_{d,k} \right) \nonumber\\
	&\quad
	=
	\sum_{i=1}^{N}\left(\mybold{h}_{\text{Obs},k}^i - \mybold{\mu}_{k}^{i}\right)^H \mybold{\gamma}_{k}^{i} \left(\mybold{h}_{\text{Obs},k}^i - \mybold{\mu_i}\right), 
	\end{align}
	where \vspace{-0.4cm}
	\begin{align}
	& \mybold{\mu}_{k}^{i} = \mybold{B}_{k} \hat{\mybold{h}}_{k|k-1}^i,  \\
	&\mybold{\gamma}_{k}^i= \left(
	\mybold{B}_k
	\left[ diag \left( \mybold{p}_{k|k-1}^i\right) \right]
	\mybold{B}^{H}_{k}+
	\sigma^2_{w}\eye{Q}
	\right)^{-1}, 
	\label{eq:gamma1} \end{align} 
	in which the first term and the second term of $\mybold{\gamma}_{k}^i$ represent the estimation error covariance and the noise covariance, respectively. $diag(\mybold{x})$ denotes creating a diagonal matrix whose main diagonal given by the elements of $\mybold{x}$.
	Then, the distortion parameters can be found by: \vspace{-0.3cm}
	
	\begin{align}
	\label{eq:loglikelihood2}
	\left \langle   \hat{\Omega}_{d,k},\hat{\Omega}_{0,k}\right \rangle =\underset{{\Omega}_{d,k},{\Omega}_{0,k}}{\arg\min} \, g\left( \Omega_{d,k},{\Omega}_{0,k}\right).
	\end{align}
	Although the NLL function $g$ is non convex, it can be divided into several constraint intervals inspired by the Nyquist sampling theorem, in which $g$ is locally convex. In each of the constraint intervals, we find the local minimum by using the constrained Newton algorithm \cite{boyd2004convex}. After that the global minimum is selected from them. Details can be found in our previous work \cite{chu2019phase}.
	
	\noindent After phase distortion parameter estimation the KF process is performed to update the channel state. We use $\mybold{K}_{k}=[
	\mybold{\kappa}_{k}^{1},  \mybold{\kappa}_{k}^{2},  \ldots,  \mybold{\kappa}_{k}^{N}
	] $ to denote the Kalman gain of the MIMO channel with\vspace{-0.1cm}
	\begin{align}
	\mybold{ \kappa}_{k}^i  = &
	\left[ diag \left( \mybold{p}_{k|k-1}^i\right) \right]
	\hat{\mybold{B}}_{k}^{H} \nonumber\\ &
	\cdot \left(\hat{\mybold{B}}_{k}
	\left[ diag \left( \mybold{p}_{k|k-1}^i\right) \right]
	\hat{\mybold{B}}^{H}_{k}+
	\sigma^2_{w}\eye{Q}
	\right)^{-1},
	\end{align}
	where $\hat{\mybold{B}}_k= e^{j\hat{\Omega}_{0,k}} \mybold{E}(\hat{\Omega}_{d,k}) \mybold{C} $. 
	Then, the channel state and the error covariance matrix are updated by \vspace{-0.3cm} 
	%
	%
	\begin{align}
	\hat{\mybold{h}}_{k|k}^i=
	\hat{\mybold{h}}_{k|k-1}^i+
	\mybold{\kappa}_{k}^i
	\left(
	\mybold{h}_{\text{Obs},k}^i - \hat{\mybold{B}}_k\hat{\mybold{h}}_{k|k-1}^i
	\right),\vspace{-0.1cm}
	\end{align}
	\vspace{-0.2cm}
	\begin{align}
	\label{eq:estcov}
	\mybold{p}_{k|k}^i =
	\left(
	\eye{}-\mybold{\kappa}_{k}^i
	\mybold{B}_{k}^{H}
	\right)
	\mybold{p}_{k|k-1}^i,
	\end{align}
	where $\hat{\mybold{h}}_{k|k}^i$ and $\mybold{p}_{k|k}^i$ are the $i$-th column of $\hat{\mybold{H}}_{k|k}$ and $\hat{\mybold{P}}_{k|k} $, respectively. 
	To conclude, the proposed algorithm  mainly includes three steps: prediction, phase distortion parameters estimation and updating. These three steps are iteratively executed to recover the CSI state from the distorted observations. 
	\section{Cramér Rao Lower bound }
	\label{sec:CRLB}
	\vspace{-0.2cm}
	In this section we derive the CRLB of the phase distortion estimation in~\ref{sec:CRLB1} and the channel state estimation in ~\ref{sec:CRLB2}. \vspace{-0.2cm}
	\subsection{CRLB for phase distortion parameter estimation}
	\label{sec:CRLB1}
	\vspace{-0.2cm}
	For the MAP based estimator we define $\hat{\mybold{\Omega}}_{k}={[\hat{\Omega}_{d,k},\hat{{\Omega}}_{0,k}]}^T$. As a result the MSE of the phase distortion parameters estimation is given by\vspace{-0.2cm}
	\begin{align} 
	\mathsf{MSE}(\hat{\mybold{\Omega}}_k) =  \mathbb{E} \left [{(\hat{\mybold{\Omega}}_{k} -\mybold{\Omega}_{k})}^T {(\hat{\mybold{\Omega}}_{k}-\mybold{\Omega}_{k}) } \right ].
	\end{align}  
	\vspace{-0.1cm}
	The unbiased estimator should satisfy\vspace{-0.2cm}
	\begin{align} 
	\mathsf{MSE}(\hat{\mybold{\Omega}}_{k}) \geqslant CRLB(\mybold{\Omega}_{k}) \vspace{-0.2cm}
	.\end{align}
	The CRLB of $\hat{\mybold{\Omega}}_{k}$ is defined as~\cite{EstTheory}  \vspace{-0.2cm}
	\begin{align}CRLB(\mybold{\Omega}_{k})= Tr \left[ {I^{-1}(\mybold{\Omega}_{k})}\right] 
	,\end{align}\vspace{-0.1cm}in which $I(\mybold{\Omega}_{k})$ is the Fisher information matrix, which is given by \vspace{-0.4cm}
	\begin{align}
	\label{eq:fisher1}
	I({\mybold{\Omega}}_{k})& = 
	\begin{bmatrix}
	\mathbb{E}\left[ \frac{\partial^2  g }{\partial {{\Omega}}^2_{d,k}}\right] 	& \mathbb{E}\left[  \frac{\partial^2  g }{\partial {{\Omega}}_{d,k} {{\Omega}}_{0,k} }\right] \\ \mathbb{E}\left[ 
	\frac{\partial^2  g}{\partial {{\Omega}}_{0,k} {{\Omega}}_{d,k}  }\right] 	& \mathbb{E}\left[  \frac{\partial^2  g}{\partial {{\Omega}}^2_{0,k}}\right]
	\end{bmatrix}.
	\end{align}Here, $g$ is the NLL function according to~\eqref{eq:loglikelihood}. 
	For simplicity, to derive the CRLB of ${\mybold{\Omega}}_{k}$ we assume the channel state $\hat{\mybold{H}}_{k|k-1}$ is perfectly predicted, which means the channel estimation error covariance $\mybold{P}_{k|k-1}$ approximately equals to zero. Thus,  $\gamma_{k}^i$ in ~\eqref{eq:gamma1} simplifies to $
	{(\sigma^2_{w}\eye{Q}
		)}^{-1}.  $
	In the following derivation of this subsection, we neglect the time index $k$ for readability. Taking the second derivative of $g$ with respect to ${\Omega}_{d}$, ${\Omega}_{0}$ and the mixed term we have \vspace{-0.3cm}
	\begin{align}
	\label{eq:derivate2}
	\frac{\partial^2  g}{\partial {{\Omega}_{d}}^2} = \frac{2}{\sigma^2_{w}}\sum_{i=1}^{N} \re\left[  { e^{j\Omega_{0}} \mybold{h}_{\text{Obs}}^{iH} \mybold{Q} \circ \mybold{Q} \mybold{E}  \mybold{C} \mybold{h}^i }\right],\end{align} \vspace{-0.45cm}
	\begin{align}
	\frac{\partial^2  g}{\partial {{\Omega}_{d}{\Omega}_{0}}} = \frac{2}{\sigma^2_{w}}\sum_{i=1}^{N} \re\left[  { e^{j\Omega_{0}} \mybold{h}_{\text{Obs}}^{iH} \mybold{Q} \mybold{E}  \mybold{C} \mybold{h}^i }\right],\end{align} \vspace{-0.35cm}
	\begin{align}
	\frac{\partial^2  g}{\partial {{\Omega}_{0}}^2} = \frac{2}{\sigma^2_{w}}\sum_{i=1}^{N} \re\left[  { e^{j\Omega_{0}} \mybold{h}_{\text{Obs}}^{iH}  \mybold{E}  \mybold{C} \mybold{h}^i }\right],\vspace{-0.1cm}
	\end{align}
	where $\mybold{Q} = diag([q_1,q_2,\ldots,q_Q]$), $\circ$ denotes the element-wise multiplication.  Then we have \vspace{-0.3cm}
	\begin{align}
	\mathbb{E} \left\lbrace\frac{\partial^2  g }{\partial {{\Omega}_{d}}^2} \right\rbrace  &= \frac{2}{\sigma^2_{w}}\sum_{i=1}^{N}\mathbb{E} \left \{  \re\left[  { e^{j\Omega_{0}} \mybold{h}_{\text{Obs}}^{iH} \mybold{Q} \circ \mybold{Q} \mybold{E}  \mybold{C} \mybold{h}^i }\right]  \right \}\nonumber \end{align} \vspace{-0.35cm}
	\begin{align}
	\quad \overset{(a)}{=} \frac{2}{\sigma^2_{w}}\sum_{i=1}^{N}\mathbb{E}\left \{  \mybold{h}^{iH}  \mybold{C}^{H} \mybold{Q} \circ \mybold{Q} \mybold{C}
	\mybold{h}^{i}  \right \}
	\nonumber\end{align} \vspace{-0.35cm}
	\begin{align}
	\overset{(b)}{=}\frac{2}{\sigma^2_{w}} \sum_{i=1}^{N} Tr (\mybold{C}^{H} \mybold{Q} \circ \mybold{Q}  \mybold{C} \mybold{\Sigma}^{i}),\vspace{-0.1cm}
	\end{align}
	where $(a)$ is by substituting the observation equation~\eqref{eq:obsEq} and $(b)$ exploits the symmetric property of $\mybold{C}^{H} \mybold{Q} \circ  \mybold{Q}  \mybold{C}$, $\mybold{\Sigma}^{i} = \mathbb{E} [\mybold{h}^{i}\mybold{h}^{iH}]$ denotes the channel covariance. The remaining components of~\eqref{eq:fisher1} are derived similarly and thus skipped due to lack of space. Finally, we get the CRLB: \vspace{-0.3cm}
	\begin{align}
	\label{CRLB1}
	&CRLB(\mybold{\Omega})=  \frac{\sigma^2_{w}}{2} \cdot\\
	& Tr{\begin{bmatrix}
		\sum_{i=1}^{N} Tr (\mybold{C}^{H} \mybold{Q} \circ  \mybold{Q} \mybold{C} \mybold{\Sigma}^{i})	&\sum_{i=1}^{N}Tr (\mybold{C}^{H} \mybold{Q} \mybold{C} \mybold{\Sigma}^{i}) \\\sum_{i=1}^{N} Tr (\mybold{C}^{H} \mybold{Q} \mybold{C} \mybold{\Sigma}^{i})
		&\sum_{i=1}^{N} Tr (\mybold{C}^{H} \mybold{C} \mybold{\Sigma}^{i})
		\end{bmatrix}}^{-1} \nonumber
	\end{align}
	\vspace{-2.2cm}
	\subsection{CRLB for channel state estimation}
	\label{sec:CRLB2}
	For the KF based channel state estimator $\hat{\mybold{H}}_{k|k}$ of $\mybold{H}_{k}$ in Sec.~\ref{sec:algorithm}, we define the MSE as: \vspace{-0.2cm}
	\begin{align} 
	\vspace{-0.2cm}\mathsf{MSE}(\hat{\mybold{H}}_{k|k})=  \mathbb{E}\left\lbrace \sum_{i=1}^{N}\left[   {(\hat{\mybold{h}}_{k|k}^i -\mybold{h}_{k}^i )}^H (\hat{\mybold{h}}_{k|k}^i -\mybold{h}_{k}^i) \right]  \right\rbrace .\nonumber
	\end{align} 
	To derive the CRLB of the channel state estimation, we assume that the phase distortion parameters are accurately estimated. Thus, the measurements matrix ${\mybold{B}}_k= e^{j{\Omega}_{0,k}} \mybold{E}({\Omega}_{d,k}) \mybold{C} $ is evaluated by the true phase distortion parameters. Following~\cite{EstimationandDetectionLimits},~\cite{havlik2015performance} the filtering CRLB ${\mybold{J}}_{k|k}^i$ and the one-step prediction CRLB  ${\mybold{J}}_{k+1|k}^i$ of $i$-th channel $\mybold{h}_{k}^i$ can be recursively computed by \vspace{-0.3cm}
	\begin{align} 
	\label{CRLB2}
	&{\mybold{J}}_{k|k}^i = 
	\mybold{J}_{k|k-1}^i-\mybold{J}_{k|k-1}^i \mybold{B}_{k}^{H} \nonumber\\ &{(\mybold{B}_{k} \mybold{J}_{k|k-1}^i \mybold{B}_{k}^{H} + \sigma^2_{w}\eye{Q} )}^{-1} \mybold{B}_{k} \mybold{J}_{k|k-1}^i,
	\end{align} \vspace{-0.45cm}
	\begin{align}  
	\label{CRLB3}
	{\mybold{J}}_{k+1|k}^i = {\alpha}^2 \mybold{J}_{k|k}^i  +\sigma^{2}_{v(i,k)}\eye{L},
	\end{align}
	initialized with $\mybold{J}_{0|-1}^i =\eye{L}$. Obviously the filtering CRLB ${\mybold{J}}_{k|k}^i$ and the one-step-prediction CRLB ${\mybold{J}}_{k+1|k}^i$ are equivalent to the updated and the predicted estimation error covariance ${\mybold{p}}_{k|k}^i$ in~\eqref{eq:estcov} and ${\mybold{p}}_{k+1|k}^i$  in~\eqref{eq:ppred}, computed with the true channel statistic and true phase distortion parameters. In this paper, we apply the filtering CRLB ${\mybold{J}}_{k|k}^i$ to evaluate our algorithm. Thus, the MSE of MIMO channel estimation should satisfy: \vspace{-0.3cm}
	
	\begin{align} 
	\label{CRLB3}
	\mathsf{MSE}(\hat{\mybold{H}}_{k|k}) \geqslant {\left[  \sum_{i=1}^{N} Tr\left(  {\mybold{J}}_{k|k}^i\right) \right] }^{-1} 
	\vspace{-0.5cm}
	\end{align}
	\section{Simulation and experimental Results}
	\vspace{-0.1cm}
	\label{sec:results}
	\begin{figure*}
		\centering
		\includegraphics[width=1\linewidth]{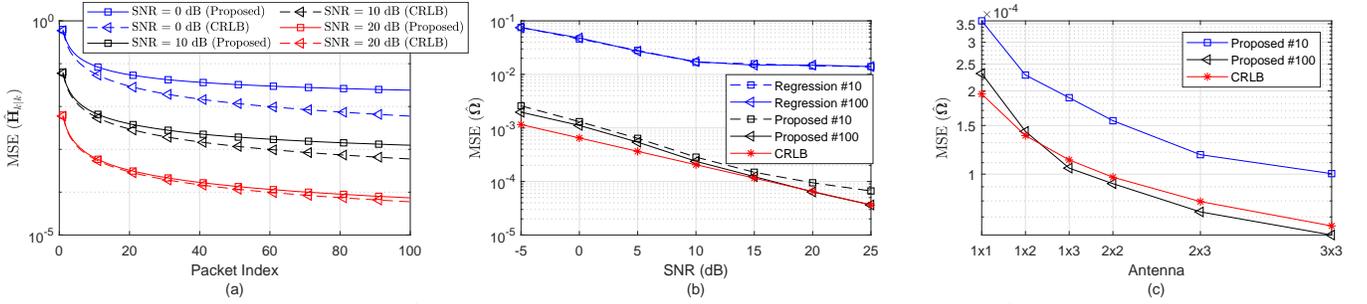} \vspace{-0.7cm}
		\caption{Simulation Results: (a) MSE$(\hat{\mybold{H}}_{k|k})$ with respect to the packet index (3x3) ; (b) MSE$(\hat{\mybold{\Omega}})$ with respect to SNR after 10 and 100 arrived packets (3x3); (c) Performance comparison of different antenna setups after 10 and 100 arrived packets (SNR = 20dB)}
		\label{fig:results} \vspace{-0.68cm}
	\end{figure*}
	
	\begin{figure}
		
		\centering
		\includegraphics[width=0.6\linewidth]{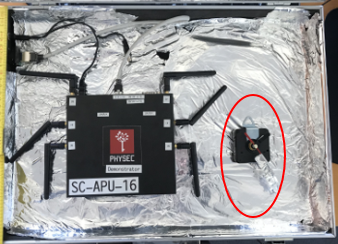}\vspace{-0.1cm}
		\caption{Experimental setup}
		\label{fig:capture} \vspace{-0.6cm}
	\end{figure}
	
	In this section, we evaluate the performance of the proposed algorithm. For the synthetic data, we generate the CSI in the time domain by a tapped-delay line model. 
	The $l$-th tap of $i$-th channel $h_{k,l}^{i}$ is modelled as a complex Gaussian random variable with zero mean and covariance ${\sigma}_{i,l}^2$ ($\sum_{l}{\sigma}_{i,l}^2 =1$).
	The channel variation is modelled using $ \alpha = { 0.5}^{10 ^{-3}}$, which means after $10 ^{3}$ packets the correlation of the channel has dropped by half. According to the IEEE 802.11n standard\cite{IEEEWiFi}, we consider a $3 \times 3$ MIMO system (and subsystems thereof) with 114 pilots. The phase slope error and offset error are randomly generated in the range of $[-0.2,0.2]$ and $[-\pi,\pi]$ at each packet, respectively. To evaluate the performance of the proposed algorithm, we ran  $10 ^ 4 $ Monte Carlo simulations and perform Kalman filtering on 100 observed CSI packets for every simulation. 

	\noindent We illustrate the simulation results in~\figref{fig:results}. From~\figref{fig:results} (a) we can see that as the index of observed CSI packets increases, the accuracy of the proposed method gradually improves. This can be attributed to the convergence property of the KF process. In addition, at high signal-to-noise ratio (SNR) the MSE of channel estimation is close to the derived CRLB in Sec.~\ref{sec:CRLB2}. At low SNR the MSE deviates from the CRLB. This is due to the fact, at low SNR the phase distortion parameters cannot be estimated accurately. In~\figref{fig:results} (b) we compare the performance of the proposed method with \textit{linear regression}~\cite{kotaru2015spotfi,ma2018signfi,ma2019wifi}. As can be seen from the figure the proposed method outperforms \textit{linear regression} significantly. Moreover, at high SNR the MSE of the proposed phase distortion parameter estimation after 100 packets matches the derived CRLB in Sec.~\ref{sec:CRLB1}. 
	The performance of different antenna setups with fixed SNR are compared in~\figref{fig:results} (c). It can be seen that even though the number of antennas of $2 \times 2$ system and $1 \times 3$ system are same, the $2 \times 2$ performs better. This is because $2  \times 2 $ has more channels than $1 \times 3$. According to~\eqref{eq:loglikelihood} and~\eqref{eq:loglikelihood2} all channel information are used to jointly estimate the phase error. That is to say the higher the number of channels, the better the performance. Note that the MSE of the phase distortion parameter estimation at packet 100 is slightly lower than the CRLB. The reason is that the MAP estimator considers the prior knowledge of the phase distortion ($ \Omega_{d,k} \in [-0.2,0.2]$ and $ \Omega_{0,k} \in [-\pi,\pi]$), while the CRLB is derived without the prior knowledge.
	\begin{figure}
		\centering
		\includegraphics[width=1\linewidth]{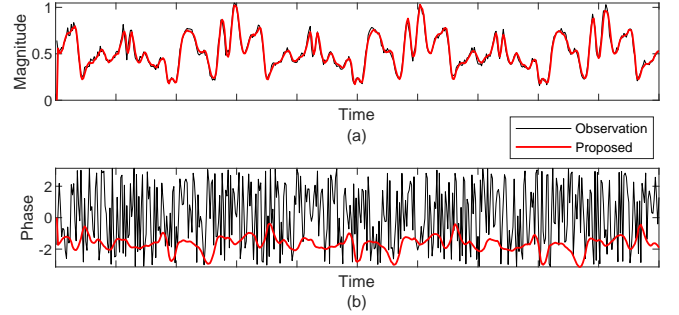} \vspace{-0.5cm}
		\caption{Experimental Results (Tx0 - Rx0, Subcarrier 1): (a) Magnitude; (b) Phase}
		\label{fig:expresults} \vspace{-0.6cm}
	\end{figure}
	
	\noindent To verify the validity of the proposed method, CSI measurements are collected using a single board computer (SBC) with 2 COTS WiFi cards in an aluminum box. Each WiFi card is equipped with 3 antennas for transmitting and receiving, respectively. A rotating reflector is placed between the antennas (marked with red circle in~\figref{fig:capture}). Due to the space limitation, here we only plot the CSI at subcarrier 1 from Tx0-Rx0, but our algorithm is actually performed at all subcarriers of the $3 \times 3$ MIMO channel. From~\figref{fig:expresults} (a), it can be easily seen that the magnitude of the measurements varies smoothly over time. Furthermore, periodic magnitude changes caused by the rotating reflector are observed. Unlike the magnitude the measured phase cf.~\figref{fig:expresults} (b) is severely impeded by random phase distortions due to the lack of time and frequency synchronization. We cannot extract any useful information from the measured phase, which makes it impossible to use in practice. However, after applying the proposed method the estimated phase changes smoothly over time and a clear periodic behavior can be observed, which means the random phase distortions are effectively removed. Thus, the experimental results nicely verify that our proposed method can track changes of the real-world channel while eliminating the random phase distortions.
	\vspace{-0.3cm}
	\section{Conclusion}
	\label{sec:conclusion}
	\vspace{-0.3cm}
	In this paper, we have proposed a phase recovery method based on KF and MAP for dynamic MIMO channels. Simulation and experimental results have demonstrated that our algorithm can effectively eliminate the phase errors and track channel changes. As future work, our findings can be applied to WiFi sensing applications, such as user authentication and activity recognition.
	\bibliographystyle{IEEEbib}
	\bibliography{refs}
\end{document}

%% file: preamble.tex
\usepackage{cite}

\usepackage{graphicx}

\graphicspath{{./figures}}
\DeclareGraphicsExtensions{.pdf}
\usepackage[cmex10]{amsmath}
\usepackage{amssymb}
\usepackage{amsthm}
\usepackage{stfloats}

\usepackage{hyperref}

\usepackage{multirow}
\usepackage{bm}
\usepackage{algorithmic}
\usepackage{xcolor}
\usepackage{tikz}
\usetikzlibrary{shapes,arrows,positioning,plotmarks,shadows,calc,matrix,fit,patterns}
\usepackage{pgfplots}
\usepackage{todonotes}
\usepackage{datetime}
\usepackage{cancel}
\usepackage{hhline}
\usepackage[overload]{empheq}
\usepackage{cases}
\usepackage{nicefrac}
\usepackage{etoolbox}
\usepackage{multirow}
\usepackage{algorithm}
\usepackage{algorithmic}
\usepackage{subcaption}
\usepackage{cleveref}

\usetikzlibrary{plotmarks}
\usetikzlibrary{arrows.meta}
\usepackage{grffile}
\usepackage{amsmath}

\usepackage{dsfont}

\newcommand{\mybold}[1]{\bm{#1}}
\newcommand{\figref}[1]{Fig.~\ref{#1}}

\newcommand{\expv}{\mathbb{E}}

\DeclareMathOperator{\re}{Re}

\newcommand{\eye}[1]{\mybold{I}_{#1}}

\newcommand{\diag}{\text{diag}}

\newtheoremstyle{remarkmod}
  {\topsep}   
  {\topsep}   
  {\normalfont}  
  {0pt}       
  {\itshape} 
  {.}         
  {5pt plus 1pt minus 1pt} 
  {}          
\theoremstyle{remarkmod}

\makeatletter
\newcommand*{\textoverline}[1]{$\overline{\hbox{#1}}\m@th$}
\makeatother

\tikzstyle{sum} = [draw, fill=blue!20, circle, node distance=1cm]
\tikzstyle{dot} = [draw, circle, minimum size=0.2pt,scale=0.3,fill=black,black]

\pgfdeclarelayer{background}
\pgfdeclarelayer{foreground}
\pgfsetlayers{background,main,foreground}

\newcounter{hints}
\renewcommand{\thehints}{\alph{hints}}
\newcommand{\hintedrel}[2][]{%
  \stepcounter{hints}%
  \if\relax\detokenize{#1}\relax\else\csxdef{hint@#1}{\thehints}\fi
  \mathrel{\overset{(\thehints)}{\vphantom{\le}{#2}}}%
}

\makeatletter
\newcommand*{\rom}[1]{\expandafter\@slowromancap\romannumeral #1@}
\makeatother

\makeatletter
\newcommand{\ALC@comblock}[1]{\ifthenelse{\equal{#1}{default}}%
{}{\textbf{#1}}}
\newenvironment{ALC@bl}{\begin{ALC@g}}{\end{ALC@g}}
\newcommand{\BLOCK}[2][default]{
	\ALC@it\ALC@comblock{#1}\ #2\begin{ALC@bl}
}
\newcommand{\ENDBLOCK}{
	\end{ALC@bl}
}

\makeatother

\newtoggle{showproof}
\toggletrue{showproof}